\documentclass{article}

\usepackage{PRIMEarxiv}

\usepackage[utf8]{inputenc} 
\usepackage[T1]{fontenc}    
\usepackage{hyperref}       
\usepackage{url}            
\usepackage{booktabs}       
\usepackage{amsfonts}       
\usepackage{nicefrac}       
\usepackage{microtype}      
\usepackage{lipsum}
\usepackage{changepage}
\usepackage{fancyhdr}       
\usepackage{graphicx}       
\graphicspath{{media/}}     

\title{Temporal Analysis of Misinformation on Parler 
}

\author{
  Eliana Norton\\
School of Journalism and Communication\\
The University of Oregon\\
  \texttt{elianan@uoregon.edu}\\
   \And
  Thaïs Thomas \\
College of Arts and Sciences\\
The University of Oregon\\
  \texttt{thaist@uoregon.edu} \\
  \And
   Akaash Kolluri\\
  Computational Media Lab\\
  The University of Texas at Austin\\
  \texttt{akaashrkolluri@computationalmedialab.com} \\
  \And
  Torie Hyunsik Kim \\
  School of Journalism and Media \\
  The University of Texas at Austin \\
  \texttt{torie.kim@utexas.edu} \\
  \And
   Dhiraj Murthy \\
School of Journalism and Media\\
   The University of Texas at Austin \\
   \texttt{Dhiraj.Murthy@austin.utexas.edu} \\
}

\begin{document}
\maketitle

\begin{abstract}
Social media platforms have facilitated the rapid spread of dis- and mis-information. Parler, a US-based fringe social media platform that positions itself as a champion of free-speech, has had substantial information integrity issues. In this study, we seek to characterize temporal misinformation trends on Parler. Comparing a dataset of 189 million posts and comments from Parler against 1591 rated claims (false, barely true, half true, mostly true, pants on fire, true) from Politifact, we identified 231,881 accuracy-labeled posts on Parler. We used BERT-Topic to thematically analyze the Politifact claims, and then compared trends in these categories to real world events to contextualize their distribution. We identified three distinct categories of misinformation circulating on Parler: COVID-19, the 2020 presidential election, and the Black Lives Matter movement. Our results are significant, with a surprising 69.2\% of posts in our dataset found to be 'false' and 7.6\% 'barely true'. We also found that when Parler posts ('parleys') containing misinformation were posted increased around major events (e.g., George Floyd's murder).
\end{abstract}


\section{Introduction}
Social media platforms have been linked to viral misinformation dissemination  \cite{muhammed}. While many social media platforms have taken steps to respond and prevent harms from misinformation \cite{Twitter, Facebook}, Parler, a US-based alt-right conservative social media platform, Parler intentionally limits fact-checking on its platform in the name of free speech \cite{brown_2021}. Moreover, Parler posts were found to be spreaders of harmful  misinformation \cite{yurieff_fung_o'sullivan_2021}. Furthermore, posts removed from mainstream social media platforms due to information integrity systems have remained on Parler \cite{pressman_2021}. For example, misinformation stating that Donald Trump won the 2020 US election and that Democrats vote rigged spread widely on Parler but were labeled as misinformation on other main-stream social media platforms \cite{newhouse_2020}. COVID-19 misinformation was also widely spread on Parler \cite{sardarizadeh_2020}.

Through a case study of Parler which identifies types of misinformation, their distribution, and their relation to real-world events, this study add to our understanding of the role of niche, extreme social media platforms and the roles they might play in the circulation of misinformation online. Misinformation has been widely studied on mainstream platforms \cite{ferreira_caceres_sosa_lawrence_sestacovschi_tidd-johnson_rasool_gadamidi_ozair_pandav_cuevas-lou_et, kolluri_murthy_2021}. While studies have characterized Parler as a platform with a large volume of posts with misinformation related to the 2020 election \cite{hitkul_prabhu_guhathakurta_jain_subramanian_reddy_sehgal_karandikar_gulati_arora_et} and anti-vaccine misinformation \cite{baines_ittefaq_abwao_2021}, there remains a dearth in the literature of those that empirically quantify misinformation dissemination across the platform. 

\section{Background}
\subsection{The Emergence of Alt-right Social Media Platforms and Parler}

As mainstream social media platforms such as Facebook, Twitter, Instagram, and even YouTube, started to regulate content, alternative right-wing platforms such as Parler gained popularity among users in the US who were censored or banned from mainstream platforms \cite{otala_kurtic_grasso_liu_matthews_madraki_2021}. While many studies examine misinformation on mainstream social media platforms such as Twitter, there are far fewer studies about misinformation that focus on alternative, far-right platforms such as Parler \cite{matamoros-fernández_farkas_2021}. It is critical to study a variety of non-mainstream social media platforms, such as Parler in order to gain valuable insights that studies of Twitter cannot provide. Parler functions very similarly to Twitter. Posts on Parler are called ‘parleys’ and can be ‘echoed’, which is similar to retweeting on Twitter. Individual users are called ‘citizens’ and can be ‘verified’ by presenting a valid identification (e.g., submitting a photograph of themselves and a government-valid photo identification - \cite{aliapoulios_bevensee_blackburn_bradlyn_de}. In the same manner as ‘likes’ work on Facebook, Parleys can only be upvoted; however, comments on parleys can be either upvoted or downvoted.

Parler is a predominantly US-focused, right-wing social networking service launched in 2018, which gained popularity by advertising minimal rules and content guidelines, drawing attention from users censored or banned on other social media platforms for violating their terms of service \cite{hitkul_prabhu_guhathakurta_jain_subramanian_reddy_sehgal_karandikar_gulati_arora_et}. The move to Parler was led by a group of far-right opinion leaders actively encouraging migration away from Twitter, such as Mark Levin, a famous talk radio host, and Maria Bartiromo, Fox News's news anchor \cite{isaac_browning_2020}. This exodus was generally comprised of “conservatives, Trump supporters, religious, and patriot individuals” \cite{aliapoulios_bevensee_blackburn_bradlyn_de} who were attracted by the minimal censorship of the platform. Those included some of the same groups found to be responsible for instigating the U.S. Capitol Insurrection, which led to hosting services such as Amazon to de-platform Parler due to its role in giving individuals a space to organize the event \cite{Agarwal_Ananthakrishnan_Tucker_2022}. Before Parler was de-platformed, it had over 10 total million users, with, at its peak on November 9th, 1.4 million users joining in a single day \cite{aliapoulios_bevensee_blackburn_bradlyn_de}. This sudden surge of individuals on Parler was largely due to real-life events such as the American presidential election \cite{aliapoulios_bevensee_blackburn_bradlyn_de}. 

The hashtags mentioned in Parler posts tend to be skewed to the US right wing (e.g., \#stopthesteal, \#maga, \#trump2020, \#votefraud) \cite{hitkul_prabhu_guhathakurta_jain_subramanian_reddy_sehgal_karandikar_gulati_arora_et}. Unlike Twitter where trending hashtags are more evenly distributed, Parler had a skewed distribution of trending hashtags eg. \#stopthesteal was present in 70\% of parley- \cite{hitkul_prabhu_guhathakurta_jain_subramanian_reddy_sehgal_karandikar_gulati_arora_et}. This relative homogeneity of political opinions among Parler users (i.e., right-leaning) contributes to the medium’s role as an echo chamber. This is similar to Gab, which is also driven by a small number of elite users, termed “super participants” \cite{zhou_dredze_broniatowski_adler_2019}. \cite{pieroni_jachim_jachim_sharevski_2021} comment that this “turns the platform into an echo chamber for its largest users.” A few right-wing leaders make up a high percentage of posts on Parler, and the traffic is not organic, according to \cite{nimmo_2019} calculation of identifying traffic manipulation on social media \cite{hitkul_prabhu_guhathakurta_jain_subramanian_reddy_sehgal_karandikar_gulati_arora_et}. 
 
Previous work has found  that right-wing social media activity has been found to engender subsequent, off-line right-wing events. Using geo-temporal data from Parler alongside ‘right-wing unrest’ data, which is represented by the Armed Conflict Location and Event Data (ACLED) database, \cite{karell_linke_holland_hendrickson_2021} discovered that prior social media activity and neighboring counties’ right-wing events lead to subsequent right-wing unrest in the county. They also found that this does not hold with the inverse, meaning that right wing unrest does not inspire right wing social media activity. Moreover, others have found that there is a strong positive correlation between war fatalities in foreign countries and the number of Parler video uploads from a county \cite{mcalexander_rubin_williams_2021}. \cite{mcalexander_rubin_williams_2021} also identified the correlation between the percentage of the county’s population being white, the number of people serving in the military before 2003, and Parler activity.

\subsection{Misinformation on Parler}

Misinformation is false or inaccurate information, which is not necessarily intentionally propagated \cite{american}. Mainstream platforms such as Twitter and Reddit actively monitor their content and attach warnings to information identified as potentially containing misinformation. For example, Twitter banned former US President Donald Trump from the social media platform \cite{conger_isaac_2021}. As it is advised by media literacy authorities for users to critically evaluate information sources \cite{metzger_flanagin_zwarun_2003}, users on the prominent platforms often actively flag potential misinformation themselves \cite{achimescu_chachev_2020}. In contrast, Parler does not have such a system to curb the spread of misinformation and problematic posts appear and spread on the platform constantly \cite{yurieff_fung_o'sullivan_2021}. Parler’s prime coincided with the start of the COVID-19 pandemic and the 2020 US presidential election, which enabled it to play a significant role in the propagation of misinformation related to these events. For example, in the case of the 2020 US presidential election, Parler provided an important space for like-minded individuals to post about election fraud-related conspiracies and other election-related misinformation under the hashtags ‘\#stopthesteal’ or ‘\#trump2020’ \cite{hitkul_prabhu_guhathakurta_jain_subramanian_reddy_sehgal_karandikar_gulati_arora_et}. Specifically, misinformation about electoral fraud online decreased by 73\% shortly after Twitter suspended Trump’s account and Parler was de-platformed \cite{rupar_2021}. Parler been found to amplify misinformation and disinformation for right-leaning Americans on issues such as COVID-19 \cite{sardarizadeh_2020}. For example, posts on Parler were found to have supported anti-vaccination (anti-vax) campaigns as well as disseminating misinformation and disinformation, such as vaccines being used as a population control mechanism \cite{baines_ittefaq_abwao_2021}.

\subsection{Impact of Misinformation}
Despite the efforts taken by mainstream platforms to mitigate the spread of misinformation \cite{ferreira_caceres_sosa_lawrence_sestacovschi_tidd-johnson_rasool_gadamidi_ozair_pandav_cuevas-lou_et}, there remains a plethora of both accurate and inaccurate information, making it difficult to rapidly gauge the veracity of content on social media \cite{donovan_wardle_starbird_zadrozny_2020}. Unlike research produced by experts, misinformation is dangerous because it needs no evidence base, knowledge or methods to produce \cite{donovan_wardle_starbird_zadrozny_2020}. Especially during the COVID-19 pandemic, misinformation and conspiracies in the pandemic have blurred the lines between normally separate groups, causing further tension in racialized posts as they become intertwined with scientific disputes \cite{donovan_wardle_starbird_zadrozny_2020}. For example, before the pandemic, anti-vaccine and neonazi groups would have been distinct, but collective sentiments against “liberal” scientific issues united these groups and has disproportionately disadvantaged Black Americans \cite{Poteat_Millett_Nelson_Beyrer_2020}. Due to the mass polarization that seems to be fueled by misinformation, some have termed the coronavirus pandemic as an infodemic \cite{donovan_wardle_starbird_zadrozny_2020}. 

With the spread of misinformation, scientific claims have become increasingly politically polarized, and as a result, some people distrust science if it goes against their beliefs \cite{oreskes_2019}. There has been much misinformation revolving around cures for COVID-19, from essential oils to cow excreta. However none of the so-called cures have been backed up by scientifically proven evidence \cite{ferreira_caceres_sosa_lawrence_sestacovschi_tidd-johnson_rasool_gadamidi_ozair_pandav_cuevas-lou_et}. COVID-19 misinformation is often dangerous, as demonstrated in June 2020, when 15 people in the US consumed disinfectant because they believed it would prevent them from catching COVID-19, leading to four people dying and three people becoming permanently visually impaired \cite{ferreira_caceres_sosa_lawrence_sestacovschi_tidd-johnson_rasool_gadamidi_ozair_pandav_cuevas-lou_et}.

Misinformation reinforces cognitive biases, further leading to racism and polarized political and social beliefs \cite{UCI}. The chain of misinformation makes it so that those believing an initial piece of misinformation can continue to seek out similar information, thus creating entire communities and mindsets around a false occurrence \cite{Del}. Due to the algorithms that platforms are built on, users are continuously presented with new content, leading to more and more opportunities for misinformation to become the dominant content \cite{YouTube_2022}. Media has a certain way of reinforcing what the user wants to believe, which, in many cases, involves the negative stereotyping of groups said user does not particularly like. Racial stereotyping, in particular, is one of the most commonly accepted yet detrimental forms of racism and misinformation present in the media. In the United States, people of color watch film and television with the underlying notion that the media aims to help maintain white supremacy \cite{hooks_1992}. In a hegemonic white society such as ours, these stereotypes single out minority groups and maintain white power \cite{seiter_1986}.

\subsection{Detection of Misinformation}
Social media platforms flag posts as “rated false” or “disputed,” but  these do not tend to change people’s views \cite{clayton_blair_busam_forstner_glance_green_kawata_kovvuri_martin_morgan_et}. That being said, fact checking remains critical both to misinformation detection as well as helping to better inform the public. Politifact, a political fact-checking website, uses a 6-point scale to show the authenticity of claims made by elected officials \cite{cui_wang_lee_2019}. Pennsylvania State University created a fact-checking framework, called SAME, which uses Politifact and GossipCop, making it more effective than other state-of-the-art methods \cite{cui_wang_lee_2019}. Politifact was found to have an accuracy rate of 93\%, a specificity rate of 90\%, a precision rate of 97\%, and a recall value of 90\%, with its overall rating being 90\% \cite{dixit_bhagat_dangi_2022}. Fact-checking websites like Politifact  help to improve the quality of information that is reaching society \cite{cui_wang_lee_2019}. Politifact uses fact-based methods to determine the accuracy of news and has been used to validate and determine the effectiveness of fact-checking models \cite{kaliyar_goswami_narang_2021, Braun_Van, Shu_Mahudeswaran_Wang_Lee_Liu_2018}.

Machine learning has also provided automated and scalable systems for detecting misinformation. Machine learning is a branch of AI that uses data to predict human interaction. This is the same thing that is used for chatbots and Netflix algorithms \cite{brown_2021}. It also efficiently uses algorithms to predict misinformation. Aside from Politifact, fuzzy matching is another machine learning tactic that is used to identify similar elements of text. There have been other studies using similar methods of fuzzy matching and Politifact data to better detect misinformation \cite{Jiang_Baumgartner_Ittycheriah_Yu_2020, Rani_Das_Bhardwaj_2021} However, while machine learning is more efficient, users are less likely to trust machine learning labels \cite{seo_xiong_lee_2019} than human-derived warnings on posts \cite{zhao_da_yan_2021}.

\section{Research Questions}

Given the dearth of studies specifically quantifying the amount of misinformation on niche, extreme social media platforms, this study seeks to answer the following research questions: RQ1: What types of misinformation circulate on Parler?, RQ2: How has the distribution of different types of true and false information changed over time on Parler? and RQ3: Is the frequency of posts with misinformation associated with real world events? 

\section{Methods}
\subsection{Datasets}

We used a publicly available dataset of 183 million comments and posts made by 4.4 million Parler users \cite{aliapoulios_bevensee_blackburn_bradlyn_de}. The dataset is composed of 98.5 million posts and 84.5 million comments. \cite{aliapoulios_bevensee_blackburn_bradlyn_de}. We restricted our study’s scope to posts and comments made from January to September of 2020.

For our fact checked claims list, we scraped claims made between January 2020 and September 2020 from Politifact\footnote{https://www.politifact.com}, an authoritative fact-checking website that classifies a wide range of information circulated on social media into six categories (pants-on-fire, false, mostly-false, halt-true, mostly-true, true) depending on trustworthiness. There were a total of 1591 claims we collected. These instances serve as benchmarks for which we compare the information in our data source to verify as true or false.

\subsection{Fuzzy Matching}
	
We used the package FuzzyWuzzy\footnote{https://pypi.org/project/fuzzywuzzy/} to conduct an approximate string matching between the Parler post and the given misinformation key from PolitiFact. A partial ratio was calculated rather than direct string matching because the Parler posts were considerably longer than the misinformation statements. All data that had below 20\% score from FuzzyWuzzy (indicating a strong match) was kept. 

\subsection{Analytical Methods}

To answer RQ1: "What types of misinformation circulate on Parler?", we used BERTopic, a topic modeling method that uses Bidirectional Encoder Representations from Transformers (BERT) embeddings and c-TF-IDF, to thematically break the Politifact statements into categories. Since BERT produced many different categories, we manually sorted these categories into posts related to COVID-19, US Politics, and Black Lives Matter (BLM): the three most prevalent groups. 

To answer RQ2: "How has the distribution of different types of true and false information changed over time on Parler?", we created and studied graphs of the distribution of the 6 nominal categories from PolitiFact (“true”, “mostly-true”, “half-true”, “barely-true”, “false”,”pants-on-fire”) and their distribution over each month from January to September 2020.

To answer RQ3: “Is the frequency of posts with misinformation associated with real world events?”, we first created comprehensive timelines of key events related to our 3 categories – COVID-19, the US election, and BLM. For COVID-19, we used a timeline from the Center for Disease Control that discusses key milestones in US cases numbers as well as political responses to COVID-19, the role out of vaccines, and economic milestones \cite{centers}. For BLM, we used an existing timeline of milestone \cite{taylor_2020, ap, wikipedia_2023c, wikipedia_2023d, abraham_2020}. For US election/politics, we used a timeline of key events in the 2020 presidential election \cite{wikipedia_2023a, wikipedia_2023b, congressional}. See Appendix A for tables with timelines for all categories.

After collecting data on the temporal distribution of accurate and inaccurate information related to our categories (organized monthly), we analyzed the relationship between temporal misinformation trends by month and our timeline of key events.

\section{Results}
\subsection{Categories of Misinformation}
To answer RQ1: “What types of misinformation circulate on Parler?", we successfully categorized content into three main categories: COVID-19, the 2020 US presidential election, and the Black Lives Matter (BLM) Movement. Table 1 illustrates the distribution of themes by category based on our BERTopic results. The top 15 categories are included.  The majority of topics are  related to politics. In this category, misinformation related to voting (1), Joe Biden (2), Nancy Pelosi (4), taxes (7), the conflict between Iraq and Iran (8), Kamala Harris (9), and schools (10) were found. Under COVID-19, we found misinformation related to face masks (3), Bill Gates and Anthony Fauci (5), deaths due to COVID-19 (6), COVID-19 (13), the relationship between COVID and China (14), and water killing the COVID-19 virus (15). Under BLM, misinformation related to the movement (11) and Jacob Blake (12) were mostly found.

 \begin{center}

\begin{tabular}{ |c|c|p{9.8cm}| }

 \hline
 Category Number & Number in Category & Top Words   \\ \hline & &
 \\[-0.25cm] 1 & 100 & ballots, voting, election, mail \\[0.05cm]
2 & 68 & joe, biden, video, shows\\[0.05cm]
3 & 54 &masks, mask, wearing, face\\[0.05cm]
4 & 51 &pelosi, nancy, pelosis, speaker\\[0.05cm]
5&
 31&gates, bill, fauci, anthony\\[0.05cm]
6&
29 &covid19, death, covid, deaths\\[0.05cm]
7&
28 &tax, joe, biden, taxes\\[0.05cm]
8&
28 &iraq, iran, beirut, iranian\\[0.05cm]
9&
26 &harris, kamala, said, next\\[0.05cm]
10&
25 &teachers, schools, education, we\\[0.05cm]
11&
24 &black, lives, matter, white\\[0.05cm]
12&
22 & kenosha, jacob, police, kyle\\[0.05cm]
13&
22& coronavirus, corona, predicted, virus\\[0.05cm]
14&
21&chinese, china, coronavirus, wuhan\\[0.05cm]
15&
20& kills, coronavirus, it, water\\[0.05cm]

 \hline

\end{tabular}
  {Table 1: BERT-Topic categories of Politifact Statements count}
\end{center}

We found high levels of misinformation related to voting in the 2020 presidential election. This includes Trump’s claims of election rigging based on ballots (1) being “found” and supposedly defective machinery \cite{yen_swenson_seitz_2020}. 

During the COVID-19 pandemic, we found misinformation circulating that claimed face masks cause harm to the people wearing them \cite{ayers_chu_zhu_leas_smith_dredze_broniatowski_2021}. Initially, the CDC had attempted to be more reserved with mandating face masks in an attempt to ensure healthcare workers had adequate supplies of masks. However, they later changed this to mandate masks after learning more about the virus, sparking distrust amongst some members of the public \cite {netburn_2021}. Misinformation relating to Bill Gates and Anthony Fauci (5) tended to claim that both played a role in the origin of COVID-19, while other people claim that Bill Gates put microchips inside of the vaccine \cite{huddleston_2021}. 

BLM-related misinformation (11) tended to be related to the right-wing counter–response movements ‘all lives matter’ and ‘white lives matter’ \cite{kelly_2020}. Jacob Blake-related misinformation (15) tended to be related to Jacob Blake, who was shot in August 2020 and partially paralyzed by a police officer. As no charges were brought upon the officer by the US Justice Department, this sparked widespread protests to the defund the police \cite{barrett_2021} . Users on Parler circulated misinformation regarding Blake and the officer that shot him.

\subsection{Temporal Distribution of Misinformation}

To answer RQ2: “How has the distribution of different types of true and false information changed over time on Parler?”, we studied the monthly distribution of the varying degrees of truth found in Parler misinformation. As seen in Figure 1, the majority of claims matching Politifact on Parler were ‘strictly false’, often spiking after events that Parler users might consider outrageous. The distributions for information related to COVID-19, the US election, and BLM (the main categories of misinformation found using the BERTopic) are shown in figures 2-4 respectively. Misinformation occurs throughout the year in all three categories. As Figures 1 and 4 indicate, the peak of misinformation frequency for BLM and misinformation generally for Parler was in June and July. Meanwhile as Figure 2 illustrates, the peak of COVID-19 misinformation was seen in July. As Figure 3 indicates, the peak for US election misinformation was in July and September.

\begin{figure*}[h]
\includegraphics[width=16.5cm, height=10cm]{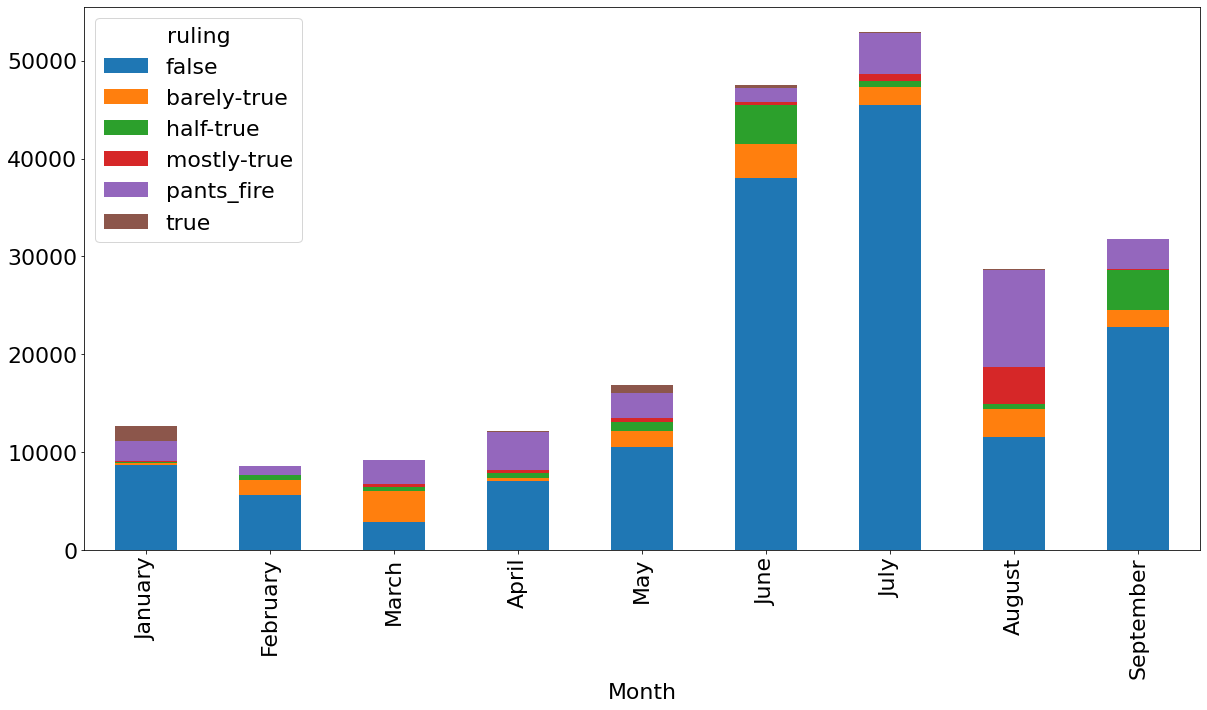}
\centering{Figure 1: The distribution of accuracy-rated posts from January to September 2020 on Parler.}
\end{figure*}

\begin{figure*}[!b]
\includegraphics[width=16.5cm, height=10cm]{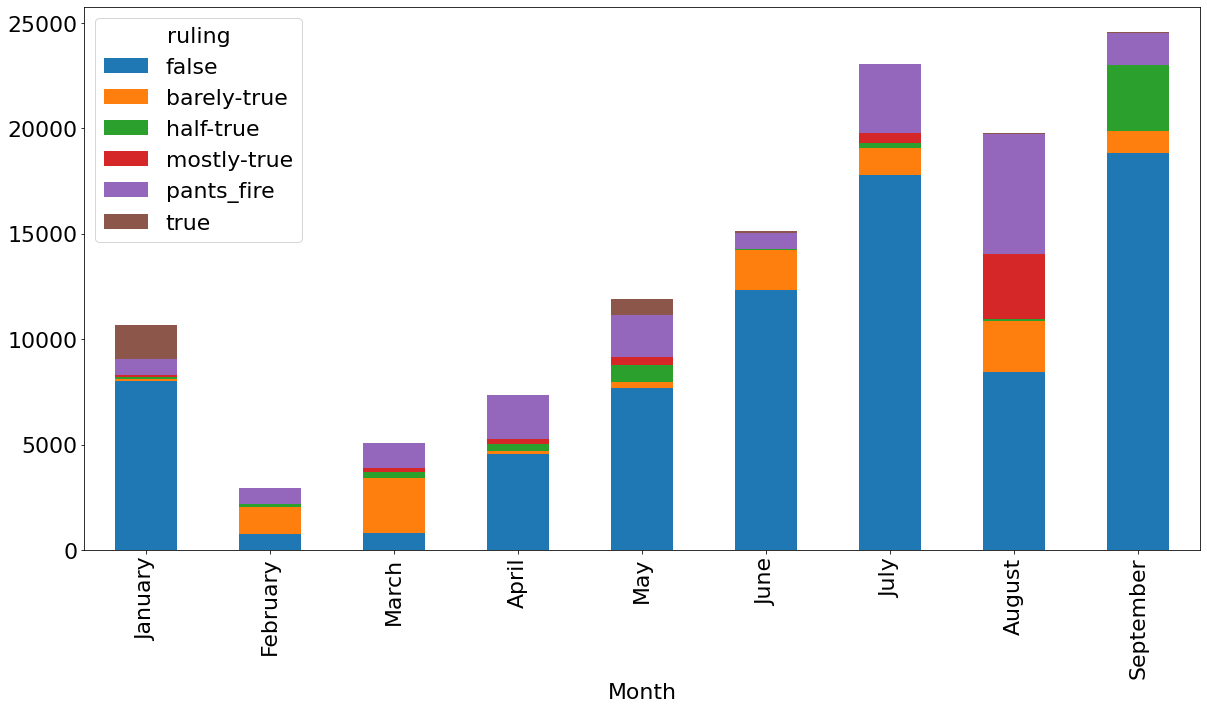}
\centering{Figure 2: The distribution of accuracy-rated posts related to US Politics from January to September 2020 on Parler.
}
\end{figure*}

\begin{figure}[h]
\includegraphics[width=16.5cm, height=10cm]{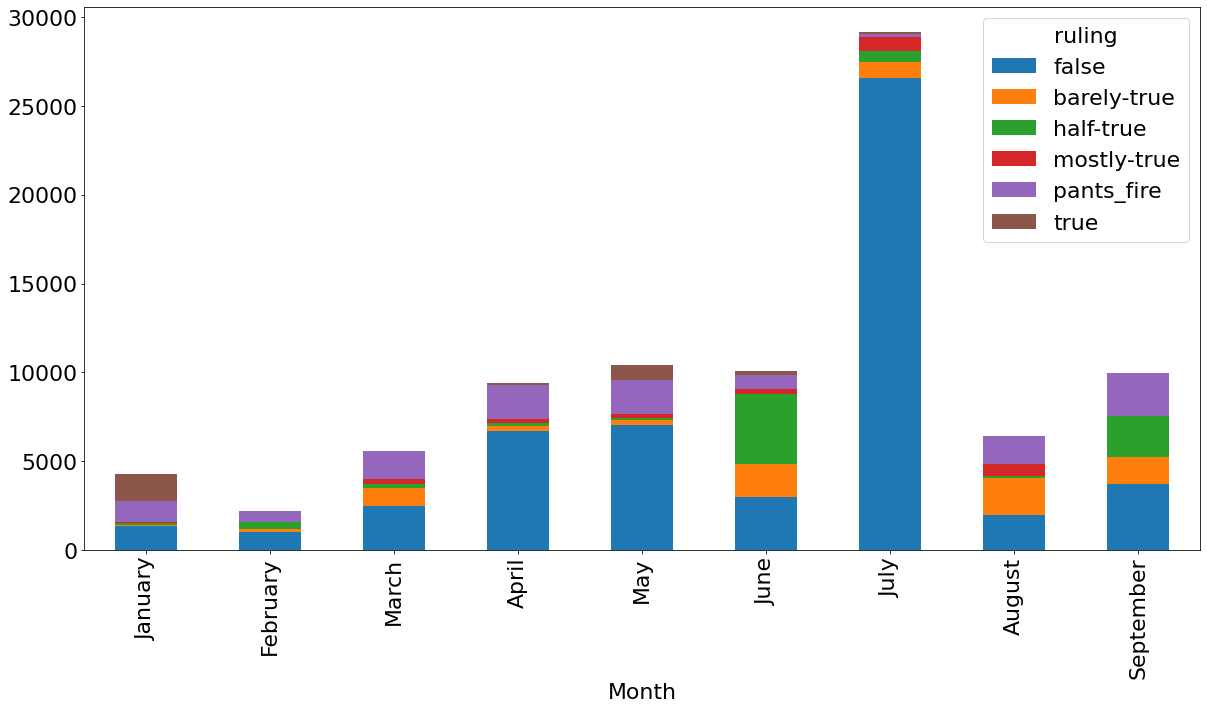}
\centering{Figure 3: The distribution of accuracy-rated posts related to COVID-19 from January to September 2020 on Parler.}
\end{figure}

\begin{figure}[!b]
\includegraphics[width=16.5cm, height=10cm]{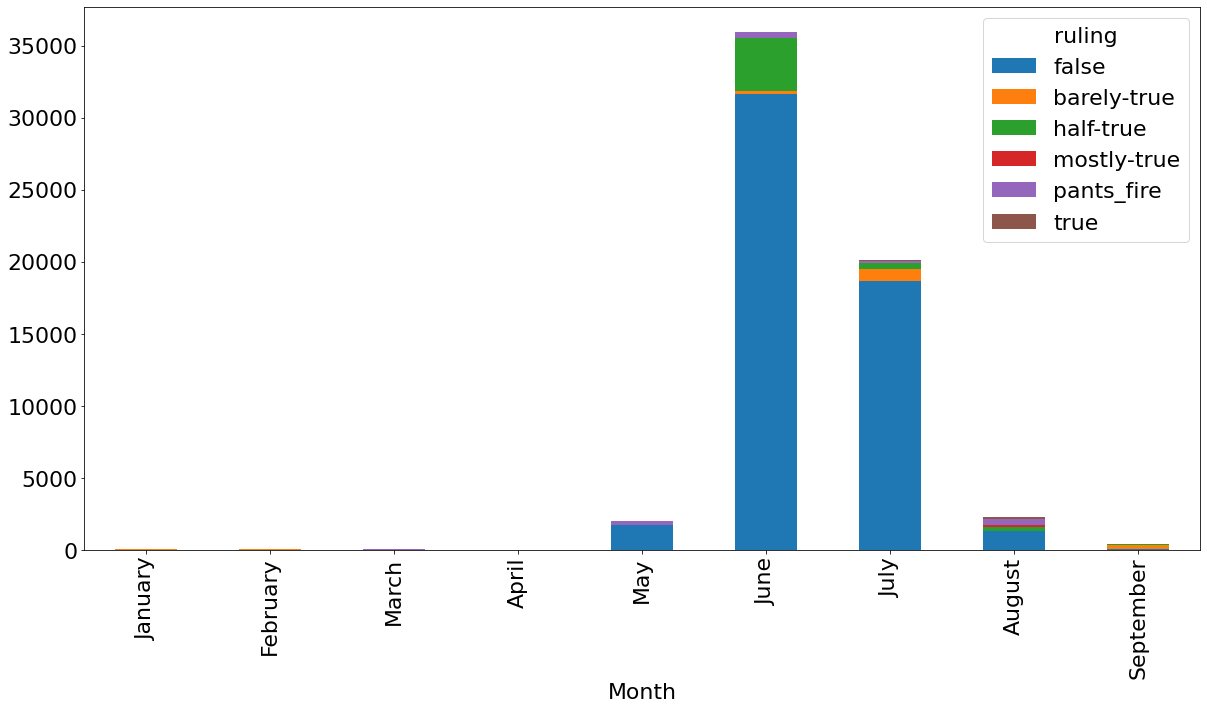}
\centering{Figure 4: The distribution of accuracy-rated posts related to the Black Lives Matter movement from January to September 2020 on Parler.}
\end{figure}

\subsection{Misinformation and Real-World Events}

To answer “RQ3: Is the frequency of posts with misinformation associated with real world events?”, we compare the trends found in figures 1-4 with the timeline of events created in Appendix A. As mentioned above, spikes in the misinformation trends occurred during more controversial, polarizing real-world events, which led to increased volumes of posts with misinformation on Parler in response.

\subsubsection{2020 US Presidential Election}
In January 2020, the US House of Representatives appointed impeachment managers, and John Delaney dropped out of the Democratic presidential race. A decrease in information about elections is seen on Parler in February, 2020 when the US Senate ruled that Donald Trump was not guilty. We found a slight increase in misinformation in March through May, 2020, with May, 2020 showing the highest levels of true information being shared. In March, 2020, Pete Buttigieg and Amy Klobuchar dropped out of the presidential race, and in April, 2020, Bernie Sanders announced that he would be endorsing Joe Biden. In May, 2020, Trump announced his plans to label Antifa as a terrorist organization and put the blame on “far-left extremist” groups for starting riots. There was a significant increase in misinformation between June and September, 2020, with September reaching a peak. In June, 2020, a poll done by CNN showed Biden in the lead by 14 points and Trump's presidential campaign asked for it to be taken down. Additionally, Biden declined to do another televised debate with Trump. Biden was officially chosen as the Democratic presidential nominee and Kamala Harris as his vice presidential running mate in August, 2020. In September, 2020, when the peak of misinformation is seen, Trump not only declined a peaceful transfer of power but he also held a rally despite concerns about COVID-19, which led to canceling a previous rally.

\subsubsection{COVID-19}
The identification of COVID-19 was met with a lot of misinformation posted on Parler. A minor increase in half-true information was seen on the app in February, 2020. However, in March and April, 2020, there were two major spikes of misinformation. In March, 2020, this was due to COVID-19 being officially declared as a pandemic by the CDC and the placement of the travel ban. In April, 2020, there was a push by some doctors and civil rights groups to make the CDC and the government release the race and ethnicity data related to COVID-19 cases so that people could quantify how the virus impacts people of color. In June 2020, the number of confirmed COVID-19 cases in the United States exceeded 2 million. This news led to an increase in fear in some individuals, which was associated with marked increases in ‘false’ and ‘barely-true’ information since the previous month. The increase in false information was associated with the announcement that asymptomatic individuals can spread the virus, the Trump Administration telling hospitals to send their hospitalization rates and equipment availability to a system used by a private contractor instead of the CDC, and the announcement CDC made asking people to wear face masks. There is a significant decrease in information about COVID-19 being shared on the platform in subsequent months.

\subsubsection{Black Lives Matter}
From the New York Times and AP News timelines, posts began emerging after the murder of George Floyd on May 25, 2020 and spiked in the following months as the trials proceeded and protests became widespread. This aligns with the frequency of misinformation on Parler. The murder of George Floyd sparked the beginning of national news coverage, coinciding with the smattering of posts beginning to appear in the BLM timeline. In June, 2020, nearly 30,000 posts related to BLM were made, which occurred synchronously with hundreds of protests occurring around the country and the Defund the Police movement becoming popular. By July, 2020, posts created dropped by about 10,000, still vastly higher than in the months prior, as the legal battle involved with the case proceeded. By August and September, 2020, the posting rate decreased significantly, dropping first to around 5,000 and 7,000, which is unsurprising as protests diminished.

\section{Limitations} 

There are limitations involved with our method of misinformation detection. Politifact is not an exhaustive list of claims that are made online. However, the findings of this study remain useful because, even operating at a subset of the misinformation on Parler, we can still discern general, relative trends. The fuzzy matching algorithm also creates inaccuracy in detection, since some claims have been misflagged and some may have been missed. Future work in these areas is needed to address these limitations. 

\section{Discussion} 
Based on our findings, we were able to successfully group the misinformation that circulated on Parler in 2020 into three distinct categories: COVID-19, US Politics, and the Black Lives Matter movement. Between the categories of Black Lives Matter, COVID-19, and US Politics, the majority of the posts were found to be false. Based on previous work \cite{shin_jian_driscoll_bar_2018}, other social media platforms such as Twitter and Reddit have shown that misinformation tends to reappear and be reposted while “true rumors” do not. This coincides with the spikes in the data signifying increases in misinformation posts on Parler. These spikes correlate to real-world events that are occurring at the same time of the posts. As discussed in Section 5, there also tends to be overlaps between categories leading to a huge spike in multiple graphs during the months of June and July. These were key months for both the BLM and COVID-19 timelines, parallel to the mass posting and protests around the death of George Floyd and the spiking increase of COVID-19 cases from 2 to 3 million cases in the US \cite{centers}. 

The level and scope of misinformation on Parler is concerning and could have influenced large numbers of people. Additionally, previous work has found that once misinformation appears in a user’s social feed, it tends to persist in their feed \cite{YouTube_2022}. On Parler which purposely avoids the use of fact-checking mechanisms, misinformation was being newly generated and shared. While it has been found that many people will actively avoid ideologies that do not reflect their own \cite{oreskes_2019}, most other platforms try to minimize the consequences of this by flagging posts and removing content or adding warning labels to posts. While previous studies have found that this is not necessarily an accurate measure of misinformation \cite{clayton_blair_busam_forstner_glance_green_kawata_kovvuri_martin_morgan_et}, it is still more useful in fact-checking than not having any detection method at all.

\section{Conclusion} 

Our  findings quantify the prevalence of misinformation on Parler and makes clear that  misinformation on niche, extreme social media platforms needs to be taken seriously and empirically studied. There can still be a recognition between the difference of opinion and information that is false regarding science and political/social movements, but creating clearer guidelines for social media platforms might contribute to a more transparent environment with the premise that fact-checking should be employed (perhaps through legal enforcement). Though we did not seek to draw causal conclusions in the context of our analysis of Parler misinformation in relation to real-world events, we noted some of the interactions between major events in Black Lives Matter and the COVID-19 pandemic and how these related to the frequency of misinformation we found on Parler. The scope of our study is rather, intentionally exploratory and does not claim specific causation. However, we hope future work can build upon our insights and study other polarized, niche social media platforms as misinformation is not just propagated by alt-right or conservative social platforms \cite{kaufman_2019}.

\section*{Acknowledgments}

This work was supported by Good Systems, a research Grand Challenge at the University of Texas at Austin. This work was supported in part by Oracle Cloud credits and related resources provided by the Oracle for Research program. We thank Kami Vinton for her thoughtful feedback on our manuscript.

\bibliographystyle{unsrt}
\bibliography{references.bib}


\newpage
 \section*{Appendix A}

\subsection*{COVID-19 Timeline of Events }
 \begin{center}
\begin{tabular}{ |c|p{14cm}| } 

 \hline
 Date & Event    \\ \hline

 & 
 \\[-0.2cm] December 2019 & First known case\\[0.1cm]
 January 2020 &
When COVID was first identified as the cause \\[0.1cm]
February 2020 & 
WHO announces the official name of the disease \\[0.1cm] March 2020 &
COVID declared a pandemic + Trump issues a travel ban \\[0.1cm]
April 2020 & Shortages in personal protective equipment + CDC announces new mask wearing girdles and recommends wearing masks when outside of home + announces government impersonation fraud (phone scams and phishing attacks that impersonate the CDC and ask for donations) + CDC begins summarizing weekly data on COVID hospitalizations, death and testing + some doctors and civil rights group tell the CDC and the government that they want race and ethnicity data related to COVID to be released so people can see the true impact of the virus on communities of color + Georgia, Alaska, and Oklahoma partially reopen their states despite it being too early \\[0.1cm]
May 2020 & Unemployment rate in US is 14.7\% + death toll surpasses 100,000  \\
June 2020 & Confirmed cases in US surpasses 2 million \\[0.1cm]
July 2020 & Confirmed casses in US surpasses 3 million + Who announces that SARS-CoV-2 virus is transmitted through the air and that asymptomatic people can spread it + CDC tells people to wear face masks “a critical tool in the fight against COVID-19” + the Trump Administration tells hospitals across the nation to stop sending critical information about COVID-19 hospitalization rates and equipment availability to the CDC and send it instead to a new system that uses a private contractor \\[0.1cm]
August 2020 & Becomes 3rd leading cause of death in US (1,000 per day + 5.4 million nationwide) + first documented case of reinfection confirmed in Hong Kong and US \\[0.1cm]
September 2020 & Death toll in US surpasses 200,000 + 1 million worldwide -  \\[0.1cm]
October 2020 & Donald Trump gets Covid \\[0.1cm]
November 2020 & Moderna’s vaccine found to be 95.4\% effective + Pfizer being 95\% \\[0.1cm]
December 2020 & First COVID vaccine granted regulatory approval + recorded death toll in US surpasses 300,000 + first case of “Alpha” variant in the US \\[0.1cm]

 \hline
 
\end{tabular}
\centering{COVID-19 Timeline of Key Events by Month \cite{centers}}
\end{center}

\newpage
\subsection*{US-Politics Timeline of Events}
 \begin{center}
\begin{tabular}{ |c|p{14cm}| } 

 \hline
 Date & Event    \\ \hline

 & 
 \\[-0.2cm]
 January 2020 & House of Reps appoint impeachment managers + Delaney drops out of the Democratic race \\[0.1cm]
February 2020 & Senate votes to acquit Trump + Walsh drops out of the Republican primary \\[0.1cm]
March 2020 & Buttigieg and Klobuchar drop out of the race and endorse Democrats + Super Tuesday \\[0.1cm]
April 2020 & Sanders endorses Biden \\[0.1cm]
May 2020 & Trump blames “far-left extremist” groups for organizing violent riots and says he plans to designate Antifa as a terrorist organization\\[0.1cm]
June 2020 & Police and national guard force protesters to clear out + Trump’s presidential campaign demands that CNN removes their poll that shows Biden leading by 14 points + Biden refuses to do a fourth debate with Trump + Biden announces that he won’t hold anymore campaign rallies due to COVID \\[0.1cm]
July 2020 & Kanye West announces that he will run for president but doesn’t officially do it + Trump states his intent to withdraw the US from the WHO + Trump struggling according to polls due to his responses to COVID and George Floyd + Biden and Trump campaigns are preparing for if the election is contested \\[0.1cm]
August 2020 & Biden officially chosen as the presidential nominee by delegates at the Democratic National Convention + Trump files a lawsuit to prevent Nevada from doing primarily mail-in voting + Biden officially chooses Kamala Harris as his vice presidential running mate \\[0.1cm]
September 2020 & Trump holds a rally after one was canceled due to COVID concerns + Ruth Bader Ginsburg passes away + Trump calls Biden the "dumbest of all candidates ... You can't have this guy as your president ... maybe I'll sign an executive order that you cannot have him as your president" + Trump declines to commit to a peaceful transfer of power after the election \\[0.1cm]
October 2020 & Melania Trump gets covid + Trump experiencing mild COVID symptoms + Trump says in an interview that Harris is a “monster” and a “communist” because of her stance on open borders + Trump refuses to participate in a virtual debate + National Woman’s March occurs \\[0.1cm]

November 2020 & “Trump Train” caravans block traffic + Trumps campaigns files a lawsuit in Michigan to stop the vote counting + Trump calls Georgia’s hand ballot recount a waste of time + Trump claims that the election was rigged which is why Biden won \\[0.1cm]
December 2020 & Electors meet to formally vote for the president and vice president, Biden won + Trump tweets an article stating that people are upset that McConnell congratulated Biden \\[0.1cm]

 \hline
 
\end{tabular}
\centering{US-Politics Timeline of Key Events by Month \cite{wikipedia_2023a, wikipedia_2023b, congressional}}
\end{center}

\newpage
\subsection*{Black Lives Matter Timeline of Events }
 \begin{center}
\begin{tabular}{ |c|p{14cm}| } 

 \hline
 Date & Event    \\ \hline

 & 
 \\[-0.2cm]
February 23, 2020 & Ahmaud Arbery is murdered. Footage begins circulating May 5\\[0.1cm]
March 13, 2020 & Breonna Taylor is murdered\\[0.1cm]
May 25, 2020 & George Floyd is murdered \\[0.1cm]
May 26, 2020 &  Bystander footage of Floyd’s murder is release shortly after police statement claiming his death as a “medical incident.” 4 officers fired + Protests in Minneapolis\\[0.1cm]
May 27, 2020 &  “Mayor Jacob Frey calls for criminal charges against Chauvin. Protests lead to unrest in Minneapolis, with some people looting and starting fires. Protests spread to other cities.”\\[0.1cm]
May 28, 2020 &  Governor Tim Walz activates National Guard\\[0.1cm]
May 29, 2020 &  Chauvin is arrested and charged with third degree murder and manslaughter  + Trump Tweets “When the looting starts, the shooting starts.” Protests continue and turn violent\\[0.1cm]
May 30, 2020 &  Trump tries to walk back his Tweet + Minneapolis mayor refers to protests as “domestic terrorism” \\[0.1cm]
May 31, 2020 &  Attorney General Keith Ellison announced to lead prosecutions + Curfews ordered by mayors\\[0.1cm]
June 1, 2020 & The county ME finds cause of death as Floyd’s heart stopping while police compressed his neck + 2 autopsies reveal Floyd’s death as homicide + Trump threatens to deploy military \\[0.1cm]
June 2, 2020 &  Minnesota Department of Human Rights launches investigation into the Minneapolis PD + Protesters gather at murder site \\[0.1cm]
June 3, 2020 &  Three officers charged for not intervening\\[0.1cm]
June 4, 2020 &  Funeral held for Floyd + Buffalo officers suspended for shoving elderly protester\\[0.1cm]
June 5, 2020 &  Minneapolis bans choke holds for police\\[0.1cm]
June 6, 2020 &  Protests continue + Officers charged with assault \\[0.1cm]
June 7, 2020 &  “A majority of Minneapolis City Council members say they support dismantling the Police Department. The idea later stalls but sparks a national debate over police reform.”\\[0.1cm]
June 8, 2020 &  Thousands pay respects to Floyd in Houston. He is buried the next day\\[0.1cm]
June 10, 2020 &  Floyd’s brother testifies before the House Judiciary Committee for police accountability\\[0.1cm]
June 16, 2020 &  Trump signs executive order for better police practices\\[0.1cm]
July 15, 2020 &  Floyd’s family sues officers and Minneapolis\\[0.1cm]
July 21, 2020 &  “The Minnesota Legislature passes a broad slate of police accountability measures that includes bans on neck restraints, chokeholds and so-called warrior-style training.”\\[0.1cm]
August 23, 2020 &  Jacob Blake shot and killed, sparking protests, many in Kenosha. Two protesters killed by Kyle Rittenhouse (he would later be acquitted of all charges in Nov)\\[0.1cm]
August 26-28, 2020 &  False rumors riot from Eddie Sole Jr.’s suicide\\[0.1cm]
September 23, 2020 &  Protests erupt after jury verdict of Breonna Taylor’s case\\[0.1cm]

 \hline
 
\end{tabular}
\centering{Black Lives Matter Timeline of Key Events by Month \cite{taylor_2020, ap, wikipedia_2023c, wikipedia_2023d}}.
\end{center}

\end{document}